\documentclass[11pt,epsf]{article}
\usepackage{amsmath}
\usepackage{amsfonts}
\usepackage{amssymb}
\usepackage{graphicx}
\usepackage{color}

\newtheorem{theorem}{Theorem}

\newcommand {\dfn} {\stackrel{\Delta} {=}}
\newcommand {\exe} {\stackrel{\cdot} {=}}
\newcommand {\lexe} {\stackrel{\cdot} {\le}}
\newcommand {\gexe} {\stackrel{\cdot} {\ge}}
\newcommand {\gea} {\stackrel{\mbox{\tiny (a)}} {\ge}}
\newcommand {\geb} {\stackrel{\mbox{\tiny (b)}} {\ge}}
\newcommand {\gec} {\stackrel{\mbox{\tiny (c)}} {\ge}}
\newcommand {\ged} {\stackrel{\mbox{\tiny (d)}} {\ge}}
\newcommand {\gee} {\stackrel{\mbox{\tiny (e)}} {\ge}}
\newcommand {\gef} {\stackrel{\mbox{\tiny (f)}} {\ge}}
\newcommand{\eqa}{\stackrel{\mbox{\tiny (a)}}{=}}

\newcommand {\bx} {\mbox{\boldmath $x$}}
\newcommand {\by} {\mbox{\boldmath $y$}}
\newcommand {\bz} {\mbox{\boldmath $z$}}

\newcommand {\bE} {\mbox{\boldmath $E$}}

\newcommand {\hP} {\hat{P}}
\newcommand {\hH} {\hat{H}}
\newcommand {\hI} {\hat{I}}

\newcommand {\bX} {\mbox{\boldmath $X$}}
\newcommand {\bY} {\mbox{\boldmath $Y$}}

\newcommand{\calE}{{\cal E}}

\newcommand{\calG}{{\cal G}}

\newcommand{\calI}{{\cal I}}

\newcommand{\calN}{{\cal N}}

\newcommand{\calT}{{\cal T}}

\newcommand{\calX}{{\cal X}}
\newcommand{\calY}{{\cal Y}}

\begin{document}
\thispagestyle{empty}
\title{Noisy Guesses}
\author{Neri Merhav}
\date{}
\maketitle

\begin{center}
The Andrew \& Erna Viterbi Faculty of Electrical Engineering\\
Technion - Israel Institute of Technology \\
Technion City, Haifa 32000, ISRAEL \\
E--mail: {\tt merhav@ee.technion.ac.il}\\
\end{center}
\vspace{1.5\baselineskip}
\setlength{\baselineskip}{1.5\baselineskip}

\begin{abstract}
We consider the problem of guessing a random, finite--alphabet, secret
$n$--vector, where the guesses are transmitted via a noisy channel.
We provide a single--letter formula for the best achievable exponential growth
rate of the $\rho$--th moment of the number of guesses, as a function of $n$. This formula
exhibits a fairly clear insight concerning the penalty due to the noise. We
describe two different randomized schemes that achieve the optimal guessing exponent.
One of them is 
fully universal in the sense of being independent of
source (that governs the vector to be guessed), the channel (that corrupts the
guesses), and the moment power $\rho$. Interestingly, it turns out that, in general, the optimal
guessing exponent function exhibits a phase transition when it is examined
either as a function of the channel parameters, or as a function of $\rho$:
as long as the channel is not too distant (in a certain sense to be defined
precisely) from the identity channel (i.e., the
clean channel), or equivalently, as long $\rho$ is larger than a certain critical
value, $\rho_{\mbox{\tiny c}}$, there is no penalty at all in the guessing exponent, compared
to the case of noiseless guessing.\\

\end{abstract}

\clearpage
\section{Introduction}

Consider the problem of guessing the realization of a finite--alphabet
random vector $\bX=(X_1,\ldots,X_n)$ using a series of yes/no questions of the
form: ``Is $\bX=\bx_1$?'', ``Is $\bX=\bx$?'', and so on, until a positive
response is received. Given a distribution on $\bX$, a commonly used
performance metric for the guessing problem
is the expected number of guessing trials
required until $\bX$ is guessed correctly, or more generally, a general moment
of this number.

The quest for \emph{guessing strategies} designed
in order to minimize
the moments of the number of guesses
has several motivations and applications in
information theory and related areas. One of them, for example, is
sequential decoding, as shown by Arikan \cite{Arikan96}, who based his work on the
earlier work of Massey \cite{Massey94}, and related the asymptotic exponent of
the best achievable guessing moment to
the R\'enyi entropy. More
recent applications of the guessing problem focus on aspects of
information security,
in particular, brute--force attacks of guessing passwords or decrypting messages protected by random
keys. For example, one may submit a sequence of guessing queries in attempt
to crack passwords -- see, e.g., \cite[Introduction]{MC19} (as well as
\cite{SHBCM19} and other references
therein) for a fairly comprehensive review
on guessing and information security, as well as for some 
historical perspective of earlier research work
on the problem of guessing in general, along with its large variety of forms and extensions.

In this paper, we consider the guessing problem where the guesser (henceforth,
Bob) submits his guesses to the party that examines the guesses (henceforth,
Alice) via a noisy discrete memoryless channel (DMC). In other words, the problem is
informally defined as follows: Alice randomly
draws an $n$--vector $\bY$ from a discrete memoryless source (DMS) $P$ of a finite
alphabet $\calY$. Bob submits a sequence of guesses, $\bx_i\in\calX^n$,
$i=1,2,\ldots$, but each guess $\bx_i$ is transmitted through a DMC
$W$, defined by a matrix single--letter transition probabilities, $\{W(y|x),~x\in\calX,~y\in\calY\}$, before arriving to
Alice. Let $\bY_1,\bY_2,\ldots$, $\bY_i\in\calY^n$, $i=1,2,\ldots$, be the
corresponding noisy versions of the guesses. Alice checks the noisy guesses
sequentially, and returns an affirmative feedback to Bob upon the first
perfect match,
$\bY_i=\bY$. The questions we are studying, in this paper, are similar to those that
were studied in earlier works on the guessing problem, namely: (i) what is the minimum achievable asymptotic exponent of
$\bE\{G^\rho\}$ (i.e., the guessing exponent), where $G$ is the number of guesses until the first success,
and $\rho$ is an arbitrary given positive real? In particular, what is the
penalty caused by the channel noise in terms of the possible increase in the
guessing exponent, compared to the noiseless case of \cite{Arikan96}? (ii) how
can one achieve this
minimum guessing exponent? 

Our main result is a single--letter formula of the best achievable guessing
exponent, i.e., the exponential growth
rate of the $\rho$--th moment of the number of guesses, as a function of $n$.
In particular, we provide two equivalent expressions of the optimal guessing
exponent. One of these expressions immediately suggests an optimal (randomized) guessing
strategy. The other expression
exhibits a fairly clear insight concerning the penalty due to the noise. We
describe two different randomized schemes that achieve the optimal guessing exponent.
One of them is
fully universal in the sense of being independent of
source $P$ (that governs the vector $\bY$ to be guessed), the channel $W$ (that corrupts the
guesses), and the power $\rho$. Obviously, the existence of a randomized
guessing scheme that achieves the optimal guessing exponent implies (albeit, not
constructively) the existence of a deterministic guessing scheme, exactly like
in standard random coding arguments. Having said that, randomized schemes have
some advantages, as was discussed in \cite{MC19}.

Interestingly, it turns out that, in general,
the optimal guessing exponent function exhibits a {\it phase transition} when it is examined
either as a function of the channel parameters, $\{W(y|x)\}$, or as a function of $\rho$:
as long as the channel is not too distant (in a certain sense to be defined
precisely in the sequel) from the identity channel (i.e., the
clean channel), or equivalently, as long $\rho$ is larger than a certain
critical value, $\rho_{\mbox{\tiny c}}$, there is no penalty whatsoever in the guessing exponent, compared
to the case of noiseless guessing.

There are several motivations for studying this problem of noisy guesses.
\begin{enumerate}
\item Since Alice and Bob might be physically remote from each other,
it is conceivable that the channel that links Bob to Alice would be noisy, and
coding/decoding may not be an option in applications where Alice has no incentive to cooperate
with Bob.
\item Alice may wish to apply a jammer as a mean of defense against a
(detected) brute--force
attack conducted by Bob.
\item We wish to explore aspects of robustness of the guessing performance to errors in the guessing mechanism.
\item We believe that the results are fairly interesting and some of them are
even quite surprising, for example, the phase transitions, and the full
universality of two of the proposed guessing schemes, in $P$, $W$ and $\rho$, as described in
the previous paragraphs.
\item It enriches the variety of perspectives and the plethora of technical
analysis tools used in the guessing problem. While the guessing problem, in
its ordinary, noiseless form, has an intimate relationship to source coding, and
hence can be solved on the basis of source coding results, this is no longer
the case in the noisy setting. Indeed,
as we shall see, the analysis
techniques are very different from those of noiseless guessing
\cite{Arikan96}. 
\end{enumerate}

The outline of the remaining part of this paper is as follows. In Section
\ref{notationconventions}, we establish notation conventions.
In Section \ref{formulation}, we formally define the problem and the
objectives. In Section \ref{mainresult}, we present our main results and
discuss them. Finally, in Section \ref{proofoftheorem1}, we prove our main
theorem.

\section{Notation Conventions}
\label{notationconventions}

Throughout the paper, random variables will be denoted by capital
letters, specific values they may take will be denoted by the
corresponding lower case letters, and their alphabets
will be denoted by calligraphic letters. Random
vectors and their realizations will be denoted,
respectively, by capital letters and the corresponding lower case letters,
both in the bold face font. Their alphabets will be superscripted by their
dimensions. For example, the random vector $\bX=(X_1,\ldots,X_n)$, ($n$ --
positive integer) may take a specific vector value $\bx=(x_1,\ldots,x_n)$
in $\calX^n$, the $n$--th order Cartesian power of $\calX$, which is
the alphabet of each component of this vector.
Sources and channels will be denoted by the letters $P$, $Q$, $V$ and $W$,
subscripted by the names of the relevant random variables/vectors and their
conditionings, if applicable, following the standard notation conventions,
e.g., $Q_X$, $P_{Y|X}$, and so on. When there is no room for ambiguity, these
subscripts will be omitted.
The probability of an event $\calE$ will be denoted by $\mbox{Pr}\{\calE\}$,
and the expectation
operator will be denoted by $\bE\{\cdot\}$.
The entropy of a random variable $X$ with a generic distribution $Q_X$ 
(or $Q$, for short) will be denoted by
$H(Q)$ or $H(Q_X)$, or $H_Q(X)$, and the Kullback--Leibler divergence between two distributions, $P$ and
$Q$, on the
same alphabet, will be denoted by $D(Q\|P)$. Likewise, for a pair of random
variables $(X,Y)$, jointly distributed according to $Q$, $H_Q(X,Y)$,
$H_Q(X|Y)$, $I_Q(X;Y)$ will denote the joint entropy, the condition entropy of
$X$ given $Y$, and the mutual information between $X$ and $Y$, respectively.
Similar notation conventions will apply to other information measures,
including those that involve more than two random variables.
The weighted divergence between two conditional distributions,
$Q_{Y|X}$ and $P_{Y|X}$, with weighting $Q_X$, is defined as
\begin{equation}
D(Q_{Y|X}\|P_{Y|X}|Q_X)=\sum_{x\in\calX}Q_X(x)\sum_{y\in\calY}Q_{Y|X}(y|x)\ln\frac{Q_{Y|X}(y|x)}{P_{Y|X}(y|x)}.
\end{equation}
For two positive sequences $a_n$ and $b_n$, the notation $a_n\exe b_n$ will
stand for equality in the exponential scale, that is,
$\lim_{n\to\infty}\frac{1}{n}\log \frac{a_n}{b_n}=0$. Similarly,
$a_n\lexe b_n$ means that
$\limsup_{n\to\infty}\frac{1}{n}\log \frac{a_n}{b_n}\le 0$, and so on.
The indicator function
of an event $\calE$ will be denoted by $\calI\{E\}$. The notation $[x]_+$
will stand for $\max\{0,x\}$.

The empirical distribution of a sequence $\bx\in\calX^n$, which will be
denoted by $\hat{P}_{\bx}$, is the vector of relative frequencies
$\hat{P}_{\bx}(x)$
of each symbol $x\in\calX$ in $\bx$.
The type class of $\bx\in\calX^n$, denoted $\calT(\bx)$, is the set of all
vectors $\bx'$
with $\hat{P}_{\bx'}=\hat{P}_{\bx}$. When we wish to emphasize the
dependence of the type class on a generic empirical distribution, say, $Q_X$, we
will denote it by
$\calT(Q_X)$. Information measures associated with empirical distributions
will be denoted with `hats' and will be subscripted by the sequences from
which they are induced. For example, the entropy associated with
$\hat{P}_{\bx}$, which is the empirical entropy of $\bx$, will be denoted by
$\hat{H}_{\bx}(X)$. An alternative notation, following the conventions
described in the previous paragraph, is $H(\hP_{\bx})$.
Similar conventions will apply to the joint empirical
distribution, the joint type class, the conditional empirical distributions
and the conditional type classes associated with pairs (and multiples) of
sequences of length $n$.
The conditional type class of $\bx$ given $\by$ w.r.t.\ a conditional
distribution $Q_{X|Y}$ will be denoted by $\calT(Q_{X|Y}|\by)$.
$\hH_{\bx\by}(X,Y)$ will designate the empirical joint entropy of $\bx$
and $\by$,
$\hH_{\bx\by}(X|Y)$ will be the empirical conditional entropy,
$\hI_{\bx\by}(X;Y)$ will
denote empirical mutual information, and so on.
Also, sometimes we will use the subscript $Q$ (like in $H_Q(X|Y)$ and
$I_Q(X;Y)$) when it is understood that $Q$ is the joint empirical distribution
associated with $(\bx,\by)$.

\section{Problem Formulation}
\label{formulation}

We consider the following scenario: Alice draws a random $n$--vector,
$\bY=(Y_1,\ldots,Y_n)$, from a discrete memoryless source
(DMS), $P$, of a finite alphabet, $\calY$. 
Bob, who is unaware of the realization of $\bY$, sequentially submits to Alice
a (possibly, infinite) sequence of guesses, $\bx_1,\bx_2,\ldots$, where each $\bx_i$ is a vector
of length $n$, whose components take on values in a finite alphabet, $\calX$.
Before arriving to Alice, each guess, $\bx_i$, undergoes a discrete memoryless channel (DMC), defined by
a matrix of single--letter input--output transition probabilities, 
$W=\{W(y|x),~x\in\calX,~y\in\calY\}$. Let $\bY_1,\bY_2,\ldots$ be the
corresponding noisy versions of $\bx_1,\bx_2,\ldots$, after being corrupted by
the DMC, $W$. Alice sequentially examines the noisy guesses and she returns 
to Bob an affirmative feedback upon the first perfect match, $\bY_i=\bY$.
Clearly, the
number of guesses, $G$, until the first successful guess, is a random
variable that depends on the source vector $\bY$ and the guesses,
$\bY_1,\bY_2,\ldots.$. It is given by
\begin{equation}
G=G(\bY,\bY_1,\bY_2,\ldots)=\sum_{k=1}^\infty
k\cdot\calI\{\bY_k=\bY\}\cdot\prod_{i=1}^{k-1}[1-\calI\{\bY_i=\bY\}].
\end{equation}
For a given list of guesses, $\calG_n=\{\bx_1,\bx_2,\ldots\}$,
$\bx_i\in\calX^n$, $i=1,2,\ldots$, the $\rho$--th moment of $G$ is given by
\begin{equation}
\label{G}
\bE_{\calG_n}\{G^\rho\}=\sum_{\by\in\calY^n}P(\by)\cdot\sum_{k=1}^\infty k^\rho\cdot
W(\by|\bx_k)\cdot\prod_{i=1}^{k-1}[1-W(\by|\bx_i)],
\end{equation}
where
\begin{equation}
W(\by|\bx)=W(y_1,\ldots,y_n|x_1,\ldots,x_n)=\prod_{t=1}^nW(y_t|x_t).
\end{equation}
Randomized guessing lists (where the deterministic guesses, $\{\bx_i\}$, are replaced by
random ones, $\{\bX_i\}$) are allowed as well. In this case, eq.\ (\ref{G})
would include also an expectation w.r.t.\ the randomness of the guesses. 
For a given sequence of lists, $\calG=\calG_1,\calG_2,\ldots$, we define
\begin{eqnarray}
\calE_{\calG}^-(\rho)&=&\liminf_{n\to\infty}\frac{\ln\bE_{\calG_n}\{G^\rho\}}{n}\\
\calE_{\calG}^+(\rho)&=&\limsup_{n\to\infty}\frac{\ln\bE_{\calG_n}\{G^\rho\}}{n},
\end{eqnarray}
and
\begin{eqnarray}
\calE^-(\rho)&=&\inf_{\calG}\calE_{\calG}^-(\rho)\\
\calE^+(\rho)&=&\inf_{\calG}\calE_{\calG}^+(\rho).
\end{eqnarray}
Our objectives, in this paper, are as follows.
\begin{enumerate} 
\item To show that $\calE^-(\rho)=\calE^+(\rho)$. The distinction between
these two functions will then disappear and both of them will be
denoted by $\calE(\rho)$.
\item To find a single--letter formula for $\calE(\rho)$, which depends on the
source $P$ and the channel $W$, in addition to the moment order, $\rho$. 
To characterize the loss in the guessing performance due to the noise
(compared to the case of noiseless guessing).
\item To derive a (possibly randomized) guessing scheme that achieves $\calE(\rho)$. 
\end{enumerate}

In fact, all three objectives will essentially be accomplished in a joint
manner. We will define a certain single--letter function, $E(\rho)$, and show that
$\calE^+(\rho)\le E(\rho)\le \calE^-(\rho)$, where the first inequality is the
direct part and the second inequality is the converse part. Using the obvious
fact that $\calE^+(\rho)\ge \calE^-(\rho)$ by their definitions, all inequalities are
in fact, equalities. 

A comment about the notation is in order. When we wish to emphasize the
dependence of $E(\rho)$ on the channel $W$ as well, we may expand the notation to
$E(\rho,W)$. The second argument, $W$, may also be replaced by a certain parameter
that completely defines $W$, like the crossover probability $q$ in case of the binary
symmetric channel (BSC).

\section{Main Results}
\label{mainresult}

Let $P$, $W$, and $\rho\ge 0$ be given.
To present the main result, we first need a few definitions.
For two given distributions, $Q_X$ and $Q_Y$, defined on $\calX$ and $\calY$,
respectively, define
\begin{equation}
\Gamma(Q_X,Q_Y)=\inf\{D(\tilde{Q}_{Y|X}\|W|Q_X):~(Q_X\odot\tilde{Q}_{Y|X})_Y=Q_Y\},
\end{equation}
where the notation $(Q_X\odot\tilde{Q}_{Y|X})_Y=Q_Y$ means that the $Y$--marginal
induced by the given $Q_X$ and by $\tilde{Q}_{Y|X}$ is constrained to be the given $Q_Y$,
i.e., $\sum_{x\in\calX}Q_X(x)\tilde{Q}_{Y|X}(y|x)=Q_Y(y)$ for all $y\in\calY$.
Next, define
\begin{equation}
\Gamma(Q_Y)=\inf_{Q_X}\Gamma(Q_X,Q_Y)=\inf_{Q_{X|Y}}D(Q_{Y|X}\|W|Q_X).
\end{equation}
Finally, we define
\begin{equation}
\label{1st}
E(\rho)=\sup_{Q_Y}\bigg\{\rho[H(Q_Y)+\Gamma(Q_Y)]-D(Q_Y\|P)\bigg\}.
\end{equation}
Theorem \ref{thm1} below states that $E(\rho)$ is, in fact, a single--letter formula
for both $\calE^+(\rho)$ and $\calE^-(\rho)$. It also provides an alternative,
equivalent expression for $E(\rho)$, which is simpler to calculate in
practice.

\begin{theorem}
\label{thm1}
Let $P$, $W$, and $\rho\ge 0$ be given and consider the problem setting
defined in Section \ref{formulation}. Assume that $W$ is such that
$W_{\max}=\max_{x,y}W(y|x) < 1$. Then,
\begin{enumerate}
\item (Converse part): $\calE^-(\rho)\ge E(\rho)$.
\item (Direct part): $\calE^+(\rho)\le E(\rho)$.
\item The function $E(\rho)$ can also be expressed as follows:
\begin{equation}
\label{2nd}
E(\rho)=\ln\left(\inf_V\sum_{y\in\calY}\frac{P(y)}{\left[\sum_{x\in\calX}V(x)W(y|x)\right]^\rho}\right),
\end{equation}
where the infimum w.r.t.\ $V$ is taken across the simplex of all probability distributions over $\calX$,
i.e., over all vectors of dimension $|\calX|$ whose components are non--negative and sum to unity.
\end{enumerate}
\end{theorem}
The proof appears in Section \ref{proofoftheorem1}. The remaining part of this
section is devoted to a discussion on Theorem \ref{thm1}.\\

\noindent
{\bf 1.\ The assumption $W_{\max} < 1$.} This technical regularity condition is needed for the proof of the converse part.
At a first glance, it seems to be rather restrictive. A closer look, however, reveals that this restriction is not too severe.
Suppose that this condition is violated, namely, there exist
pairs $(x,y)$ with $W(y|x)=1$. Let $\calY'=\calY\cap\{y:~\max_xW(y|x)=1\}$.
Now, for any type $Q_Y$ whose support is $\calY'$, and for every
$\by\in\calT(Q_Y)$, we can find $\bx$ such that $W(\by|\bx)=1$, and the
problem of guessing within such a type boils down to the problem of noiseless
guessing: whenever one wishes to guess a certain $\by$, he or she should guess instead
an $\bx$ for which $W(\by|\bx)=1$. For every $Q_Y$ whose support is not
included in $\calY'$, and every $\by\in\calT(Q_Y)$, there exists at least one
letter $y_i$ for which $W\dfn \max_xW(y_i|x)< 1$ and hence
$W(\by|\bx)=\prod_{i=1}^n W(y_i|x_i) \le W < 1$, which is still acceptable for the
derivation of the converse proof (with $W$ replacing $W_{\max}^n$).\\

\begin{figure}[h!t!b!]
\centering
\includegraphics[width=8.5cm, height=8.5cm]{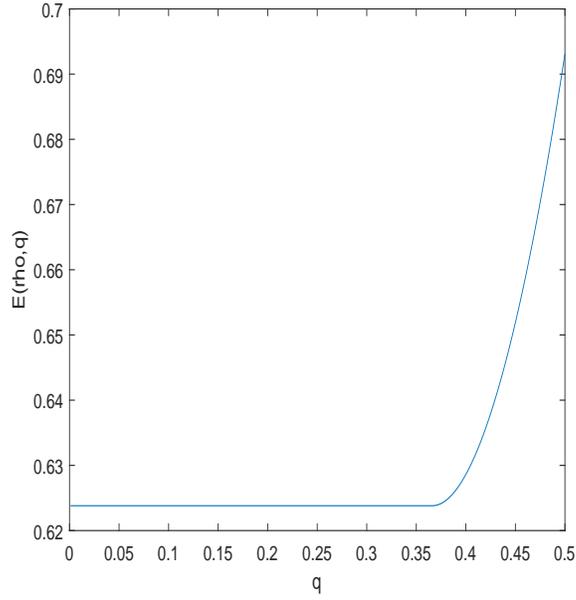}
\caption{The function $E(\rho,q)$ vs.\ $q$ for the BSS with parameter $p=0.25$ and $\rho=1$.
The critical value of $q$ is $q_c=\sqrt{p}/(\sqrt{p}+\sqrt{1-p})=0.366$.}
\label{graph1}
\end{figure}

\begin{figure}[h!t!b!]
\centering
\includegraphics[width=8.5cm, height=8.5cm]{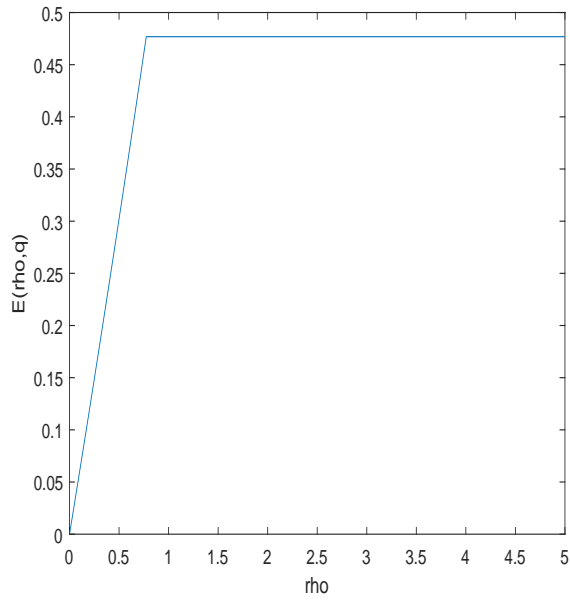}
\caption{The function $E(\rho,q)$ vs.\ $\rho$ for the BSS with parameter $p=0.25$
and the BSC with parameter $q=0.35$.
The critical value of $\rho$ is $\rho_c=\ln[(1-p)/p]/\ln[(1-q)/q]-1=0.7747$.}
\label{graph2}
\end{figure}

\noindent
{\bf 2.\ The penalty due to the noise and phase transitions.} As can be seen from 
the first expression of $E(\rho)$ (eq.\ (\ref{1st})), the term $\Gamma(Q_Y)$
expresses the unavoidable, minimum achievable penalty that Bob must suffer due to the noise. Interestingly, 
there might be situations, where there is no penalty at all as far as the guessing exponent is concerned. 
These situations can be identified even more easily from the second expression of $E(\rho)$. 
First, observe that in the noiseless case,
where $W$ is the clean channel, eq.\ (\ref{2nd}) boils down to 
$E(\rho)=\ln(\inf_Q\sum_y P(y)/[Q(y)]^\rho)$ ($Q$ being a probability distribution over $\calY$), which is achieved by
the distribution $Q^*(y)$ that is proportional to $[P(y)]^{1/(1+\rho)}$. Eq.\ (\ref{2nd}) can be 
thought of as the same minimization problem, except that now $Q$ is constrained to lie in the convex hull of
$\{W(\cdot|x),~x\in\calX\}$. 
Generally speaking, good channels have a relatively large convex hull, 
whereas bad channels have a relatively small one. In the extreme case 
of the clean channel, this convex hull is the entire simplex 
of probability distributions over $\calY$. In the other extreme, where the channel is 
completely useless, all $\{W(\cdot|x),~x\in\calX\}$ coincide with a 
single distribution over $\calY$. In this case, the convex hull of $W$ is a singleton.
Returning to the case of a general channel, $W$, 
if the above--mentioned distribution $Q^*$ happens to lie within this convex hull, then no penalty
is incurred in the guessing exponent. Interestingly, this implies that in general there might be a
{\it phase transition} in the behavior of the guessing exponent as a function of the noise level, or as a function of $\rho$.
For a given source $P$ and a given moment order $\rho$, consider a family of channels defined by a single parameter. 
For example, consider the binary case ($\calX=\calY=\{0,1\}$), where the binary source is defined by the 
parameter $p=\mbox{Pr}\{X_i=1\}$ and the family of channels are the 
binary symmetric channels (BSCs) with a crossover parameter $q\in[0,0.5]$. 
Assume also that $\rho=1$. If we increase $q$ gradually from $0$ to $0.5$, we see that as long as 
$q < q_{\mbox{\tiny c}}\dfn \sqrt{p}/(\sqrt{p}+\sqrt{1-p})$, 
the guessing exponent is the same as in the noiseless case, independently of $q$.
However, as $q$ crosses the critical value $q_c$ the guessing exponent starts to grow with $q$. The graph of the guessing exponent,
here denoted $E(\rho,q)$, is depicted in Fig.\ \ref{graph1}, for $p=0.25$ 
and $\rho=1$, where $q_{\mbox{\tiny c}}=0.366$. Likewise, it is 
possible to examine the behavior of the guessing exponent as a function of $\rho$ for fixed $q$. In the case of the BSC, 
the convex hull of $W$ is always a symmetric interval around $0.5$, i.e., the interval $[q,1-q]$. 
The optimal distribution $Q^*$, defined above is a binary distribution with probabilities,
$p^{1/(1+\rho)}/[p^{1/(1+\rho)}+(1-p)^{1/(1+\rho)}]$ and
$(1-p)^{1/(1+\rho)}/[p^{1/(1+\rho)}+(1-p)^{1/(1+\rho)}]$. As $\rho$ grows, this distribution approaches the symmetric distribution,
$(0.5,0.5)$. Therefore, there is critical value of $\rho$ beyond which this distribution falls in the convex hull of $\{q,1-q\}$.
This critical value is given by
$$\rho_{\mbox{\tiny c}}\dfn\left[\frac{\ln[(1-p)/p]}{\ln[(1-q)/q]}-1\right]_+.$$
Fig.\ \ref{graph2} depicts the guessing exponent as a function of $\rho$, with $p=0.25$ and $q=0.35$, where
$\rho_{\mbox{\tiny c}}=0.7747$.\\

\noindent
{\bf 3.\ Operative significance of the alternative expression (\ref{2nd}).}
Quite obviously, the alternative expression of $E(\rho)$, given in eq.\ (\ref{2nd}), is easier to calculate in practice, than
that of eq.\ (\ref{1st}). However, the advantage of (\ref{2nd}) goes 
considerably beyond this point, as it has a simple operative interpretation.
The denominator (without the power of $\rho$) is clearly the i.i.d.\ channel output distribution induced
by an i.i.d.\ channel input distribution $V$ and the channel $W$. If we use randomized guessing and draw all our guesses according to the memoryless source $V(\bx)=\prod_{t=1}^nV(x_t)$, then the noisy guesses will also be random guesses distributed according
to $Q(\by)=\prod_{t=1}^nQ(y_t)$, where $Q(y)=\sum_xV(x)W(y|x)$. As is well known from 
previous work (see, e.g., \cite{MC19},
\cite{SHBCM19}), randomized guessing according to an i.i.d.\ distribution $Q$ yields a guessing exponent of
$\ln(\sum_yP(y)/[Q(y)]^\rho)$. But since $Q$ is constrained to have the form $Q(y)=\sum_xV(x)W(y|x)$, 
for a given $W$, the best one can do is to minimize this expression over $V$, which is exactly what eq.\ (\ref{2nd}) tells us to do. 
It follows that the achievability of $E(\rho)$ is conceptually simple, thanks to eq.\ (\ref{2nd}): find the channel input 
distribution $V^*$ that achieves the minimum in (\ref{2nd}) and then generate random guesses according to
$V^*(\bx)=\prod_{t=1}^n V^*(x_t)$. It should be pointed out, however, that the optimal $V^*$ depends, in general, on
$P$, $W$ and $\rho$, and therefore, this achievability scheme is not universal, as it requires the knowledge of these ingredients.\\

\noindent
{\bf 4.\ A universal guessing scheme.}
While $E(\rho)$ was argued in item 3 to be achievable by a non--universal guessing scheme, it turns out that
$E(\rho)$ can also be achieved by a universal scheme that is independent of $P$, $W$ and $\rho$. 
Such a scheme is proposed in the direct part (part 2) of Theorem \ref{thm1}. This scheme is randomized: 
it is based on independent random selection of $\{\bx_i\}$ according to the same universal distribution that was proposed in
\cite{MC19}, namely,
\begin{equation}
P(\bx)=\frac{\exp\{-n\hH_{\bx}(X)\}}{\sum_{\tilde{\bx}}\exp\{-n\hH_{\tilde{\bx}}(X)\}},
\end{equation}
where $\hH_{\bx}(X)$ is the empirical entropy associated with $\bx$. The fact that this 
universal distribution continues to be asymptotically optimal even in the noisy case considered here, is not quite trivial, and it is
even fairly surprising (at least to the author). It enhances even further the powerful properties of this distribution.
We should mention also that random selection according to this distribution can be implemented efficiently in practice, as was 
shown in \cite{MC19}.\\

\noindent
{\bf 5.\ Side information.} Suppose that Bob is equipped with a side information sequence $\bz$ that is correlated with
the sequence $\by$ to be guessed. More precisely, we assume that the 
sequence pair $(\by,\bz)$ is drawn from a pair of correlated memoryless sources, 
$P(\by,\bz)=\prod_{t=1}^n P(y_t,z_t)$, but only $\bz$ is available to Bob. It turns out that our results extend straightforwardly to
this setting. Conceptually, once we condition on the given $\bz$, we are essentially back to the same setting as before. 
Thus, everything should be first conditioned on the side information, and finally, one should take the expectation w.r.t.\ the side
information. As a result, the expressions of $E(\rho)$ extend as follows:
\begin{equation}
E(\rho)=\sup_{Q_{YZ}}\{\rho[H_Q(Y|Z)+\Gamma(Q_{YZ})]-D(Q_{YZ}\|P)],
\end{equation}
where 
\begin{equation}
\Gamma(Q_{YZ})=\inf_{Q_{X|YZ}}D(Q_{Y|XZ}\|W|Q_{XZ}),
\end{equation}
and
\begin{equation}
E(\rho)=\ln\left[\sum_z P(z)\min_V\sum_{y}\frac{P(y|z)}{\left[\sum_xV(x|z)W(y|x)\right]^\rho}\right].
\end{equation}
The universal guessing distribution defined in item 4 would be replaced by 
\begin{equation}
P(\bx|\bz)=\frac{\exp\{-n\hH_{\bx\bz}(X|Z)\}}{\sum_{\tilde{\bx}}\exp\{-n\hH_{\tilde{\bx}\bz}(X|Z)\}},
\end{equation}
where $\hH_{\bx\bz}(X|Z)$ is the empirical conditional entropy of $X$ given $Z$, induced by the pair $(\bx,\bz)$ (see also
\cite{MC19}).\\

\noindent
{\bf 6.\ Sources and channels with memory.} 
Another possible extension of our results addresses sources and channels with memory. 
In particular, our setting can essentially be extended to a class of sources and
channels that obey
the following fading memory conditions for some $B \ge 0$ 
(see also \cite{AM98} and \cite{Ziv88}).
\begin{equation}
P\left\{|\ln P(y_1,\ldots,y_k|y_{-\ell},\ldots,y_0)-\ln
P(y_1,\ldots,y_k)|\ge B\right\}=0~~~\forall k,\ell\in
\calN
\end{equation}
and
\begin{eqnarray}
& &W\bigg\{|\ln
W(y_1,\ldots,y_k|x_{-\ell},\ldots,x_k,y_{-\ell},\ldots,y_0)-\nonumber\\
& &\ln W(y_1,\ldots,y_k|x_1,\ldots,x_k)|\ge B\bigg|x_{-\ell},\ldots,x_k
\bigg\}=0~~~\forall k,\ell\in\calN
\end{eqnarray}
Such sources and channels (like Markov sources and channels under certain
conditions) can be approximated by block--i.i.d.\ probability measures and
then the same derivations as before apply w.r.t.\ the super-alphabets of blocks. While one cannot expect
single--letter formulas in this case, the
point is that the universal probability distribution that is associated with
the empirical entropy associated with these blocks can be replaced (and slightly improved)
by the universal probability distribution,
$P(\bx) \propto 2^{-LZ(\bx)}$, where $LZ(\bx)$ is the length (in bits) of the
compressed version of $\bx$ by the Lempel--Ziv (LZ78) algorithm \cite{ZL78}, as was shown in
\cite{MC19}. Therefore, this universal distribution is asymptotically optimal
in terms of the guessing exponent whenever the source and the channel have
fading memory in the above defined sense.

\section{Proof of Theorem \ref{thm1}}
\label{proofoftheorem1}

\subsection{Proof of Part 1 -- the Converse Part} 

We begin from a simple preparatory step: let $\calT(Q_Y)$ be a given type
class of $\by$--vectors, and let $\bx\in\calT(Q_X)$ be given. Then,
\begin{eqnarray}
W[\calT(Q_Y)|\bx]&=&\sum_{\by\in\calT(Q_Y)}W(\by|\bx)\nonumber\\
&=&\sum_{\calT(Q_{Y|X}|\bx)\subseteq\calT(Q_Y)}|\calT(Q_{Y|X}|\bx)|\cdot\exp\{-n[H_Q(Y|X)+D(Q_{Y|X}\|W|Q_X)]\}\nonumber\\
&\exe&\sum_{\calT(Q_{Y|X}|\bx)\subseteq\calT(Q_Y)}\exp\{nH_Q(Y|X)\}\cdot\exp\{-n[H_Q(Y|X)+D(Q_{Y|X}\|W|Q_X)]\}\nonumber\\
&\exe&\exp\left\{-n\inf_{\{Q_{Y|X}:~(Q_X\odot
Q_{Y|X})_Y=Q_Y\}}D(Q_{Y|X}\|W|Q_X)\right\}\nonumber\\
&=&\exp\{-n\Gamma(Q_X,Q_Y)\},
\end{eqnarray}
and so,
\begin{equation}
\max_{\bx\in\calX^n}W[\calT(Q_Y)|\bx]\exe\exp\{-n\inf_{Q_X}\Gamma(Q_X,Q_Y)\}=\exp\{-n\Gamma(Q_Y)\}.
\end{equation}
Now, let $c > 0$ be an arbitrary constant, independent
of $n$ (say, $c=1$), and for a given $Q_Y$, define
\begin{equation}
k(Q_Y)=\bigg\lceil
\frac{c\cdot|\calT(Q_Y)|}{\max_{\bx\in\calX^n}W[\calT(Q_Y)|\bx]}\bigg\rceil\exe
\exp\{n[H_Q(Y)+\Gamma(Q_Y)]\}.
\end{equation}
We next derive a lower bound to $\bE\{G^\rho|\bY\in\calT(Q_Y)\}$ for a given
$\calT(Q_Y)$.
\begin{eqnarray}
\bE\{G^\rho|\bY\in\calT(Q_Y)\}&\gea&[k(Q_Y)]^\rho\cdot\mbox{Pr}\{G\ge
k(Q_Y)|\bY\in\calT(Q_Y)\}\nonumber\\
&=&[k(Q_Y)]^\rho\cdot\frac{1}{|\calT(Q_Y)|}\sum_{\by\in\calT(Q_Y)}\prod_{i=1}^{k(Q_Y)}[1-W(\by|\bx_i)]\nonumber\\
&=&[k(Q_Y)]^\rho\cdot\frac{1}{|\calT(Q_Y)|}\sum_{\by\in\calT(Q_Y)}
\exp\left\{\sum_{i=1}^{k(Q_Y)}\ln[1-W(\by|\bx_i)]\right\}\nonumber\\
&\geb&[k(Q_Y)]^\rho\frac{1}{|\calT(Q_Y)|}
\sum_{\by\in\calT(Q_Y)}\exp\left\{-\sum_{i=1}^{k(Q_Y)}\frac{W(\by|\bx_i)}{1-W(\by|\bx_i)}\right\}\nonumber\\
&\gec&[k(Q_Y)]^\rho\cdot\frac{1}{|\calT(Q_Y)|}\sum_{\by\in\calT(Q_Y)}
\exp\left\{-\frac{1}{1-W_{\max}^n}\cdot\sum_{i=1}^{k(Q_Y)}W(\by|\bx_i)\right\}\nonumber\\
&\ged&[k(Q_Y)]^\rho\exp\left\{-\frac{1}{1-W_{\max}^n}\cdot\frac{1}{|\calT(Q_Y)|}\sum_{i=1}^{k(Q_Y)}\sum_{\by\in\calT(Q_Y)}
W(\by|\bx_i)\right\}\nonumber\\
&=&[k(Q_Y)]^\rho\exp\left\{-\frac{1}{(1-W_{\max}^n)|\calT(Q_Y)|}
\sum_{i=1}^{k(Q_Y)}W[\calT(Q_Y)|\bx_i]\right\}\nonumber\\
&\ge&[k(Q_Y)]^\rho\exp\left\{-\frac{k(Q_Y)}{(1-W_{\max}^n)|\calT(Q_Y)|}
\max_{\bx\in\calX^n}W[\calT(Q_Y)|\bx]\right\}\nonumber\\
&\gee&[k(Q_Y)]^\rho\exp\bigg\{-\frac{1}{(1-W_{\max}^n)}\times\nonumber\\
& &\left(\frac{c\cdot|\calT(Q_Y)|}{\max_{\bx\in\calX^n}W[\calT(Q_Y)|\bx]}+1\right)
\cdot\frac{\max_{\bx\in\calX^n}W[\calT(Q_Y)|\bx]}{|\calT(Q_Y)|}\bigg\}\nonumber\\
&=&[k(Q_Y)]^\rho\exp\left\{-\frac{1}{1-W_{\max}^n}\cdot\left(c+
\frac{\max_{\bx\in\calX^n}W[\calT(Q_Y)|\bx]}{|\calT(Q_Y)|}\right)\right\}\nonumber\\
&\gef&[k(Q_Y)]^\rho\exp\left\{-\frac{c+1}{1-W_{\max}^n}\right\}\nonumber\\
&\exe&\exp\{n\rho[H_Q(Y)+\Gamma(Q_Y)]\},
\end{eqnarray}
where (a) follows from Markov's inequality, (b) stems from the chain
$$\ln(1-x)=-\ln\left(1+\frac{x}{1-x}\right)\ge-\frac{x}{1-x},~~~~~~x<1,$$
(c) is implied by the assumption that $W_{\max}=\max_{x,y}W(y|x) < 1$, 
(d) is by Jensen's inequality, applied to the exponential function
$f(x)=e^{-x}$, (e) is by the definition of $k(Q_Y)$, and (f) is because
$\max_{\bx\in\calX^n}W[\calT(Q_Y)|\bx]\le 1$ and $|\calT(Q_Y)|\ge 1$ for any
non--empty type class.
Finally,
\begin{eqnarray}
\bE\{G^\rho\}&=&\sum_{Q_Y}P[\calT(Q_Y)]\cdot\bE\{G^\rho|\bY\in\calT(Q_Y)\}\nonumber\\
&\gexe&\sum_{Q_Y}\exp\{-nD(Q_Y\|P)\}\cdot\exp\{n\rho[H_Q(Y)+\Gamma(Q_Y)]\}\nonumber\\
&\exe&\exp\{n\max_{Q_Y}(\rho[H_Q(Y)+\Gamma(Q_Y))-D(Q_Y\|P))\}\nonumber\\
&=&e^{nE(\rho)}.
\end{eqnarray}

\subsection{Proof of Part 2 -- the Direct Part}

Consider a sequence of randomized guesses, all drawn independently from the
universal distribution,
\begin{equation}
P(\bx)=\frac{\exp\{-n\hat{H}_{\bx}(X)\}}{\sum_{\bx'\in\calX^n}\exp\{-n\hat{H}_{\bx'}(X)\}},
\end{equation}
for all $\bx\in\calX^n$.
This induces the following distribution on the $\by$--vectors:
\begin{eqnarray}
Q(\by)&\exe&\sum_{\bx\in\calX^n}\exp\{-n\hat{H}_{\bx}(X)\}\cdot
W(\by|\bx)\nonumber\\
&\exe&\sum_{\calT(Q_{X|Y}|\by)}|\calT(Q_{X|Y}|\by)|\cdot
\exp\{-n[H_Q(X)+H_Q(Y|X)+D(Q_{Y|X}\|W|Q_X)]\}\nonumber\\
&\exe&\max_{Q_{X|Y}}\exp\{n[H_Q(X|Y)-H_Q(X)+H_Q(Y|X)+D(Q_{Y|X}\|W|Q_X)]\}\nonumber\\
&\exe&\exp\{-nH_Q(Y)\}\cdot\max_{Q_{X|Y}}\exp\{-nD(Q_{Y|X}\|W|Q_X)]\}\nonumber\\
&\exe&\exp\{-n[H_Q(Y)+\Gamma(Q_Y)]\}.
\end{eqnarray}
Owing to \cite[Lemma 1]{MC19} and \cite{gis}, for any given
$\by\in\calT(Q_Y)$, the $\rho$--th moment of the
number of guesses w.r.t.\ the randomized guessing, would then be of the
exponential order of 
$1/[Q(\by)]^\rho\exe\exp\{n\rho[H_Q(Y)+\Gamma(Q_Y)]\}$. 
Finally, upon averaging this quantity with weights $P[\calT(Q_Y)]\exe
\exp\{-nD(Q_Y\|P)\}$, we clearly obtain an expression of the exponential order of
$e^{nE(\rho)}$. This completes the proof of the direct part.

\subsection{Proof of Part 3 -- the Alternative Expression}

Consider the following chain of equalities:
\begin{eqnarray}
E(\rho)&=&\sup_{Q_Y}\inf_{Q_{X|Y}}\left[\rho H(Q_Y)+\rho
D(Q_{Y|X}\|W|Q_X)-D(Q_Y\|P)\right]\nonumber\\
&=&\sup_{Q_Y}\inf_{Q_{X|Y}}\bigg[-\rho\sum_yQ_Y(y)\ln
Q_Y(y)+\rho\sum_{x,y}Q_{XY}(x,y)\ln\frac{Q_{Y|X}(y|x)}{W(y|x)}+\nonumber\\
& &\sum_yQ_Y(y)\ln\frac{P(y)}{Q_Y(y)}\bigg]\nonumber\\
&=&\sup_{Q_Y}\inf_{Q_{X|Y}}\bigg[-\rho\sum_yQ_Y(y)\ln
Q_Y(y)+\rho\sum_{x,y}Q_{XY}(x,y)\ln\frac{Q_{Y}(y)Q_{X|Y}(x|y)}{W(y|x)Q_X(x)}+\nonumber\\
& &\sum_yQ_Y(y)\ln\frac{P(y)}{Q_Y(y)}\bigg]\nonumber\\
&=&\sup_{Q_Y}\inf_{Q_{X|Y}}\bigg[
\rho\sum_{x,y}Q_{XY}(x,y)\ln\frac{Q_{X|Y}(x|y)}{W(y|x)Q_X(x)}+
\sum_yQ_Y(y)\ln\frac{P(y)}{Q_Y(y)}\bigg]\nonumber\\
&=&\sup_{Q_Y}\inf_{Q_{X|Y}}\inf_V\left[
\rho\sum_{x,y}Q_{XY}(x,y)\ln\frac{Q_{X|Y}(x|y)}{W(y|x)V(x)}+
\sum_yQ_Y(y)\ln\frac{P(y)}{Q_Y(y)}\right]\nonumber\\
&=&\sup_{Q_Y}\inf_V\inf_{Q_{X|Y}}\left[
\rho\sum_{x,y}Q_{XY}(x,y)\ln\frac{Q_{X|Y}(x|y)}{W(y|x)V(x)}+
\sum_yQ_Y(y)\ln\frac{P(y)}{Q_Y(y)}\right]\nonumber\\
&=&\sup_{Q_Y}\inf_V\left[
-\rho\sum_{y}Q_{Y}(y)\ln\left(\sum_xV(x)W(y|x)\right)+
\sum_yQ_Y(y)\ln\frac{P(y)}{Q_Y(y)}\right]\nonumber\\
&\eqa&\inf_V\sup_{Q_Y}\left[
-\rho\sum_{y}Q_{Y}(y)\ln\left(\sum_xV(x)W(y|x)\right)+
\sum_yQ_Y(y)\ln\frac{P(y)}{Q_Y(y)}\right]\nonumber\\
&=&\inf_V\ln\left(\sum_y\frac{P(y)}{\left[\sum_xV(x)W(y|x)\right]^\rho}\right)\nonumber\\
&=&\ln\left(\inf_V\sum_y\frac{P(y)}{\left[\sum_xV(x)W(y|x)\right]^\rho}\right),
\end{eqnarray}
where the equality (a) is due to the convexity in $V$ and the concavity in
$Q_Y$ of the objective function.
This completes the proof of Theorem \ref{thm1}.


\end{document}